# A Feature Dataset of Microservices-based Systems

Weipan Yang[1][23S030135@stu.hit.edu.cn], Yongchao Xing[2][22B903085@stu.hit.edu.cn],
Yiming lü[3][2201110523@stu.hit.edu.cn], Zhihao Liang[4][2201110520@stu.hit.edu.cn],
Zhiying Tu*[5][tzy_hit@hit.edu.cn]

[1-5] School of Computer Science and Technology, Harbin Institute of Technology Weihai, 264209, China

**Abstract.** Microservice architecture has become a dominant architectural style in the service-oriented software industry. Poor practices in the design and development of microservices are called microservice bad smells. In microservice bad smells research, the detection of these bad smells relies on feature data from microservices. However, there is a lack of an appropriate open-source microservice feature dataset. The availability of such datasets may contribute to the detection of microservice bad smells unexpectedly. To address this research gap, this paper collects a number of open-source microservice systems utilizing Spring Cloud. Additionally, feature metrics are established based on the architecture and interactions of Spring Boot style microservices. And an extraction program is developed. The program is then applied to the collected open-source microservice systems, extracting the necessary information, and undergoing manual verification to create an open-source feature dataset specific to microservice systems using Spring Cloud. The dataset is made available through a CSV file. We believe that both the extraction program and the dataset have the potential to contribute to the study of microservice bad smells.

## 1 Introduction

Microservice architecture has become a dominant architectural style in the service-oriented software industry [1]. Microservice architecture achieves the decoupling of a complex business into multiple small-grained microservices. Each microservice operates in its own process, allowing independent deployment, scalability, and testing, while fulfilling a specific functional responsibility. The communication between microservices relies on lightweight mechanisms [2-3].

Since the rise of microservice architectures, research on poor practices in designing and developing microservices has followed, and such research is known as microservice bad smells [4] or microservice-based antipattern research [5]. The research methodology can be broadly classified into two categories: static analysis based on source code [6-7] and analysis during system runtime [8]. The former is the primary focus of relevant research. Regrettably, most studies in this field have primarily focused on establishing a set of metrics for detecting bad smells based on bad smells definitions.

Subsequently, they evaluate the presence of bad smells by analyzing the metrics data. However, there is a relative scarcity of studies that comprehensively analyze the structure of microservice systems, extract more comprehensive feature metrics, evaluate the granularity of microservices, their design, and the interactions between different microservices based on the architecture and interaction of microservices, and subsequently explore the occurrence of poor practices within and between microservices. Additionally, it is necessary to explore the quality attributes defined in ISO25010:2023 [9], such as system modularity and maintainability, in the context of microservices. This exploration should be based on metrics extracted from the diverse fundamental elements of microservices. To address the current gap, this paper analyzes the architecture and interactions of microservices based on Spring in various Spring Cloud style microservice systems, establishes the relevant metrics of various fundamental elements of microservices in Spring Boot style, which are based on the three-tier architecture, collects open-source microservice systems, implements the extraction program1, and constructs a dataset containing microservice feature data, which in turn paves the way for exploring poor practices within and between different microservices through machine learning, heuristic algorithms, and other means. To achieve this objective, the paper presents and resolves the following three problems.

RQ1: How can an amount of Spring Cloud style microservice systems be collected as data sources and organized into a catalog for constructing feature dataset?

RQ2: How to identify the various basic elements that need to be extracted for Spring Cloud style microservice systems and extract them accordingly?

RQ3: How can the accuracy of the extracted data be validated to create a reliable dataset?

This paper presents the following studies and solutions in response to the three questions.

1. For RQ1, Spring Cloud style microservice systems are initially screened on GitHub by applying specific search conditions. Third-party libraries, frameworks, and development tools like Low-Code development platforms are excluded. Subsequently, from the remaining open-source projects, the more mature projects are selected to compile an open-source catalog[2] of Spring Cloud style microservice systems.

2. For RQ2, we analyze Spring Boot style microservices in Spring Cloud style microservice system[3]. Focusing on the three-tier architecture of individual microservices and adhering to their respective naming conventions, we identify crucial classes and interface files within microservices. On this basis, we define fourteen metrics to capture the fundamental aspects of individual microservices and derive nine supplementary metrics from the initial fourteen. By utilizing these twenty-three metrics, an initial evaluation can be conducted on the granularity, design, and interactions between different microservices. Finally, the extraction program for the metrics is implemented with the aid of toolkits like JavaParser, JGit, and Maven-Model.

3. For RQ3, this paper verifies the extracted data manually to ensure its accuracy. A part of the verification results is presented in Table 2. Ultimately, the manually verified data is curated as the feature dataset of open-source microservice systems. The dataset is made available through a .csv file[4].

The rest of the paper is structured as follows: Section II introduces the related work of this paper, Section III introduces how to construct the catalog of microservice systems, Section IV introduces how to establish the twenty-three metrics as well as the extraction logic of the program and the important algorithms. Section V elaborates on the feature data and validates its accuracy, Section VI discusses the limitations of this research, and Section VI concludes the paper by summarizing the findings and suggesting future work based on the feature dataset and extraction program.

## 2 Related work

In the research of code bad smells, researchers have extracted relevant metrics data from various open-source projects to build different datasets [10-12]. A dataset focusing on technical debts and code bad smells is created by establishing 30 distinct software metrics [10]. This was done by analyzing 33 open-source Apache Software Foundation Java projects with the aid of third-party tools like SonarQube. Tighilt et.al.[13] attempts to detect code bad smells by machine learning.

In recent years, the research on microservice bad smells has expanded extensively, resulting in continuous extensions of the bad smells catalog [14-15]. Moreover, The related detection methods for these bad smells have been continuously updated, which roughly include static analysis [7,16] and detection based on runtime data [8]. There also have attempts to apply machine learning algorithms in detection [17-18]. An open-source dataset was formed through an analysis of the effective lines of code and dependencies in 20 open-source microservice systems [17]. Abid et.al.[18] confirm the feasibility of establishing metrics, generating association rules using machine learning, and detecting the quality of web service. This confirmation indirectly supports the feasibility of integrating metrics and machine learning for the detection of microservice bad smells. However, there is a lack of work on extracting comprehensive metrics data tailored to microservice systems, and the absence of an open-source dataset that captures microservice characteristics. This constraint impedes the application of machine learning to microservice bad smell detection and limits the scope of investigating the correlation between the design, implementation of microservice systems and their associated quality attributes. To bridge this gap, this paper concentrates on establishing the extraction of metrics for various fundamental elements of microservices based on Spring Boot three-tier architecture in Spring Cloud style microservice systems. We implement an extraction program, collect data from open-source systems, conduct extraction and manual verification, and create an open-source dataset comprising microservice system feature data. This dataset will serve as a fundamental resource for the application of machine learning algorithms and further research in the domain of microservice bad smell detection.

---

[4] https://github.com/yang66-hash/Spring-Cloud-Microservice-Dataset.git

**Table 1.** A subset of selected microservice systems.

| Name | Service number | Multiple tags | Introduction | Stars |
|---|---|---|---|---|
| apollo | 5 | Yes | Apollo is a reliable configuration management system suitable for microservice configuration management scenarios. | 28.7K |
| gpmall | 10 | No | E-commerce platform based on microservices. | 4.8k |
| mogu_blog_v2 | 7 | Yes | Microservices-based open-source blog system. | 1.5k |
| mall4cloud | 11 | Yes | Microservices-based mall system. | 5.5k |
| microservice-recruit | 7 | No | Microservices-based open-source intelligent recruitment system. | 209 |
| siam-cloud | 9 | No | Microservices-based open-source takeaway delivery system. | 27 |
| Scblogs | 5 | Yes | Microservices-based open-source campus blog system. | 318 |
| Seckillcloud | 4 | Yes | Microservices-based mall system. | 36 |
| spring-petclinic-microservices | 7 | Yes | Distributed version of the Spring Pet-Clinic Sample Application. | 1.5k |
| train-tickets | 41 | Yes | The project is a train ticket booking system based on microservice architecture which contains 41 microservices [20]. | 627 |

## 3 Microservice system selection

Ewan Tempero et al. [19] curate and make available a collection of Java projects. However, to date, there has been no research undertaken to select and organize open-source Spring Cloud style microservice systems. We select open-source projects based on microservice architecture from GitHub, the open-source code hosting platform, as the primary source for feature data extraction. These projects are microservice systems developed based on the Java Spring Cloud framework. In this paper, we curate and share a catalog comprising 55 microservice systems. Among these projects, 13 systems have multiple versions, and 14 systems have more than 1000 Stars on GitHub, which are diverse, encompassing backend management systems, e-commerce platforms, blog systems, and other types.

In the specific search and retrieval process, the following strings were used to initially screen relevant projects: "topic:microservices language:Java" and "topic:spring-cloud language:Java".

We filter 2800 microservice system projects based on the former search condition and an additional 2700 microservice projects based on the latter. Subsequently, we exclude projects that are duplicated and other projects, including third-party libraries, frameworks and Low-Code development tools. From the remaining microservice systems, we further filter and select 55 systems for data extraction purposes. The specific screening rules are as follows.

1. The microservice system should consist of four or more microservices, and the business division among microservices should be logically sound.
2. The system should have service registration and service discovery mechanisms.
3. The preferred selection is open-source systems with multiple stable versions, meaning that the associated GitHub repository should have multiple release tags.
4. The most microservices in the system should be developed on Spring Boot.

The prioritization of selecting open-source microservice systems with at least 4 microservices and multiple versions aims to ensure that the chosen systems possess a certain level of maturity in terms of project scale and development standards. As for the fourth rule, we found the most open-source microservice systems based on Spring Cloud satisfy this. Table 1 presents the top 10 most representative open-source projects that were collected and utilized for this research, along with their corresponding introductions.

## 4    Extraction strategy

### 4.1    Establishing extraction metrics

Spring Cloud is built upon the foundations of the Spring and Spring Boot frameworks. Although the microservices architecture of Spring Cloud is inherently distributed, individual microservices, which are vertically partitioned based on business logic, are commonly developed using Spring Boot and adhere to the widely adopted three-tier architecture (shown in Fig. 1). The microservice systems collected in Section Ⅲ also confirm this. The architecture primarily involves the following four aspects:

- **The presentation layer** is responsible for receiving and handling user requests, as well as presenting results to the users. In Spring Boot style microservices, this mainly refers to the controller layer, which is typically annotated with @RestController or @Controller.We establish a series of metrics including controllerNum,  serviceImplCall, and others about APIs.

- **The business logic layer** is responsible for handling the business logic and rules of the application. In Spring Boot style microservices, this mainly refers to the service layer, which coordinates and processes the business workflows. It is annotated with @Service. We establish the metric of serviceClassNum and interfaceNum.
- **The data access layer** is responsible for interacting with the database and performing data persistence and retrieval operations. In Spring Boot style microservices, this mainly refers to repositories, which encapsulate database operations such as data manipulation (CRUD). Spring Data JPA or other ORM frameworks are commonly used for data access implementation. We also add up the data access operation interfaces to interfaceNum.
- **Model** objects are Java objects used for encapsulating and transferring data. They represent business data in the microservice and are responsible for operations such as data retrieval, storage, and modification. We establish a series of entity-related metrics and Data Transfer Object metrics.

Besides the metrics mentioned above, we also establish other metrics, such as codeSize, serviceCall, serviceCalled, to show the effective lines of source code of microservices, and the invocation relationships among microservices. All these metrics enable an evaluation of the microservices' granularity, design, and interaction relationships. Detailed information regarding these metrics is presented in Table 2. Among them, fourteen metrics are directly extracted, while the remaining nine metrics are calculated and derived based on these fourteen metrics.

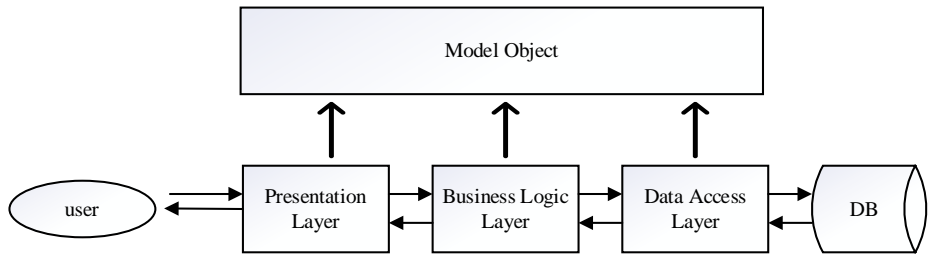

**Fig. 1.** Three-tier architecture of microservice based on Spring Boot

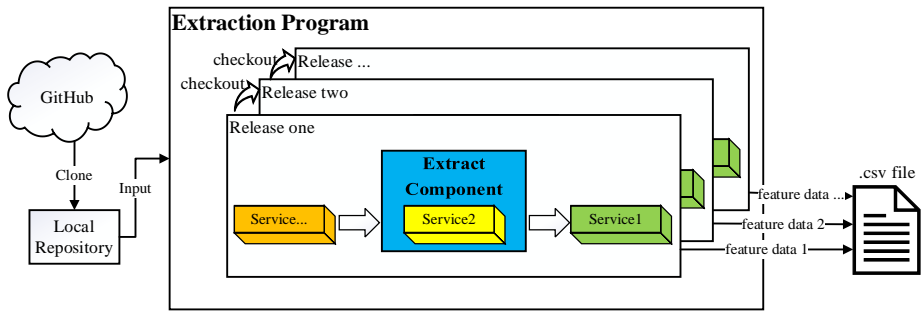

**Fig. 2.** Extraction program working framework.

### 4.2 Extraction program for metrics

The microservice systems gathered in Section III consist of Maven projects based on Spring Cloud style. We parse and extract feature data from each microservice system. The working framework for the extraction program is illustrated in Fig. 2. Firstly, the microservice system is cloned from GitHub to the local repository, and then the extraction program is executed to obtain feature data for each release of the repository. Lastly, the extracted data from each release is aggregated into a .csv file and stored in a predetermined file path.

The most important part of the extraction program is the **extract component**, which consists of the following four small parts of the extraction functionality.

**Extract code count.** Drawing inspiration from existing open-source tool cloc[5] for counting effective lines of source code.

**Extract feature data about Various classes related to Spring Boot style microservices.** The extraction of different classes primarily involves analyzing the programming conventions of the traditional three-tier architecture in Spring Boot. This analysis combines package naming conventions with regular expressions for filtering and matching, and also includes the direct identification of classes with specific annotations. The main classes comprise controller, serviceImpl, interface, Entity, DTO, and abstract. The Controller corresponds to the Presentation Layer, serviceImpl corresponds to the Business Logic Layer, interface corresponds to the Data Access Layer's data operation interfaces, Entity corresponds to the database mapping and persistence objects in the Model, and DTO corresponds to the data transfer objects between the Presentation Layer and Business Logic Layer in the Model. Among them, accurately identifying the quantity of Entity and DTO classes poses a relative challenge, whereas the remaining parts can be accurately identified based on their corresponding annotations.

- Entity classes: Entity refers to the object classes, which are utilized for mapping and persistence with the database. In this study, we perform initial filtering of Java files by applying regular expressions (1) to the package path. Subsequently, the presence of relevant annotations is determined based on the utilized data access dependencies. If these annotations are detected, the class is classified as an entity. For instance, when employing the Spring Data JPA, the presence of the @Entity annotation in the file is verified to classify it as an entity class.

$$[/\\\\](?i)(entity|pojo|model|domain|bean)[/\\\\] \tag{1}$$

- Data Transfer Object classes: Serve as data transfer objects between the Presentation Layer and Business Logic Layer. By matching class names that start or end with the string "dto" or by using regular expression (2) to match all class files within the package named "dto" in the package path.

$$[/\\\\](?i)dto[/\\\\] \tag{2}$$

---

**Feature data about APIs exposed.** Each method within the Controller classes is parsed, to identify annotations such as @RequestMapping (or its variants for Get, Post, Put, Delete, Patch) that associate HTTP requests with the "public" declaration. If these criteria are met, the method is considered a valid API and documented accordingly. Additionally, the maximum value of the parameter list size for each API is recorded, and the API version is detected by capturing the HTTP path and matching regular expression (3).

$$/(?i)v\backslash d+(\backslash.\backslash d+)?$$  (3)

**Interaction between microservices.** The extraction of this part is challenging. We aim to extract the invocation relationships among microservices in a microservice system. These invocation relationships provide insight into the system's internal interactions and assist in identifying some potential microservice bad smells. Currently, most microservices communicate with each other using either RestTemplate or Feign.

For the method of the RestTemplate class, we utilizes the JavaParser to parse methods in Java classes, aiming to determine whether the methods of the RestTemplate class are invoked for HTTP requests and to extract the first parameter representing the URL. However, when extracting the first parameter directly, the result may be a combination of a variable name or method call with part of a URL, rather than the actual URL. This may result in inaccuracy in the analysis of microservice invocations. Hence, we employ a method that recursively extracts and replaces the value of the variable or method call, effectively transforming the URL expression into a concatenated string format. And then matching the microservice name with the regular expression (4) for counting. An example is shown in Fig. 3. Firstly, the exchange method call should be detected. After that, variable station_food_service_url should be replaced by the method call of getServiceUrl, and then this method call will be replaced by "http://" + "ts-station-food-service" in recursively. Finally, the microservice name will be matched by the regular expression.

$$\backslash\backslash S*(service|Service|SERVICE)$$  (4)

**Fig. 3.** Explanation of the image depicting the detection method.

Algorithm 1 describes the process of extracting the method call of the RestTemplate class and matching regular expression (4) for the microservice name being invoked. In our implementation, we parse on each Java class, get a list of fields (with each field corresponding to a FieldDeclaration object) and methods (with each method corresponding to a MethodDeclaration object), which are the input of Algorithm 1. And then we conduct a search within the fieldDeclarations list to identify the declaration of RestTemplate objects. If such declaration is found, we iterate through each MethodDeclaration associated with that class, applying Algorithm 1 to extract the collection of expressions representing method calls of the RestTemplate class within each method. We further analyze the URL parameters of each method call statement within the expressions. Then employ a recursive approach to substitute any remaining NameExpr (variable expression in JavaParser) and MethodCallExpr (method call expression) until the URL parameter is transformed into a concatenated string format. Subsequently, put the microservice name and call times into callMap. Finally, this approach enhances the precision of static analysis when establishing microservice invocation relationships. However, it is crucial to recognize that the scope of this URL parameter parsing method is limited to variables and method calls within the confines of the Java class in question. It is essential to note that if there are references to static constants or functions external to the class, the possibility of misjudgment arises.

When detecting microservice communication using Feign, our approach involves searching for classes or interface files that are annotated with @FeignClient. From these files, we extract the respective microservice name by retrieving the value stored in the "value" or "name" field. Subsequently, we analyze the outcome of invoking these feign clients within other Java classes.

---

**Algorithm 1**: Extract service call in method.

**Input**: The set of field declarations for parsed class currently: filedDeclarations. The set of method declarations for parsed class: methodDeclarations. The position of the method being analyzed in methodDeclarations currently: i.

**Output**: The map of microservice calls, the key is the microservice name while the call times is value: callMap.

1: expressions ← ∅;
2: callMap ← ∅;
3: method ← methodDeclarations.get(i);
4: **if** get restTemplateName methodCallExpr in method **do**
5:   add methodCallExpr to expressions;
6: **end if**
7: **for** expr in expressions **do**
8:   urlExpr ← parse first argument in expr;
9:   url ← replace variables and method Calls in urlExpr in recursive;
10: serviceName ← matching url with regular expression (6);
11: callMap.put(serviceName,callMap.getOrDefault(serviceName, 0) + 1);
12:**end for**
13:**return** callMap;

# 5    Validate of extracted data

The feature extraction program was executed on 55 microservice systems, yielding a total of 1446 data points. Subsequently, after removing data from the registry center, gateway, and instances of misjudgment, a remaining set of 1180 data points was obtained. We conducted a correctness verification of each data point by cross-referencing it with the source code manually. Fig. 4 illustrates the distribution and presence of outliers for each numerical metric within this refined dataset. The median is represented by a red solid line, while the mean is denoted by a blue dashed line. Notably, the data exhibits a power-law distribution. The data distribution of each metric exhibits a notable concentration within a narrow range. For example, upon conducting statistical analysis of metric data from various classes, it becomes evident that most data points reside within the lower range. However, a considerable number of data points also exhibit outlier characteristics. This phenomenon can be attributed to the coexistence of both small-scale open-source microservice systems and larger, more intricate ones, such as the Apollo-Portal within Apollo and select microservices within the Train-Ticket system. The former is much more numerous than the latter. Although median these larger microservices are relatively scarce in number, their presence beyond the scale of most microservices businesses contributes to the existence of outliers in the final microservice dataset.

In the previous text, it was mentioned that a manual verification of the source code was conducted for each data point. To demonstrate the effectiveness and accuracy of the extraction program, we compare the metric data obtained from representative systems with the manually inspected results (details shown in Table 3), which may have errors, but very few and close to the true answers. Therefore, we consider the results of manually inspected as the correct answers. The codeSize metric, extracted based on the inspiration of cloc, yielded similar results with cloc. Therefore, it is not presented in the comparison. Additionally, only the directly extracted metrics are presented, excluding nine metrics deriving from them.

The data in Table 3 reveals that extraction errors primarily occur in entity classes associated with the data access layer and the invocation relationships among microservices. Some errors in detecting microservice invocation relationships arose due to the communication ways of a few open-source microservice systems being updated to WebClient, which is recommended by Spring (for instance, spring-petclinic-microservices). Another set of errors was identified due to the presence of references to static variables from other files when using RestTemplate or Feign. Currently, our extraction program cannot detect this part of the invocation relationships accurately. Consequently, this results in some omissions or errors during the detection process, which were corrected manually during the later stage of human inspection.

The detection errors of entityNum and entityAttributeNum in the gpmall microservice system occurred due to non-standardized package naming. The package was named "entitys", resulting in a significant number of entity classes missed. As for the error in entityAttributeNum of Scblogs, certain entity classes in some microservices extend basic classes from the common sharing project, but the attributes in base classes are missed, resulting in the loss of some attributes from the base classes. Additionally, another issue was encountered when detecting entity classes. Some entity classes are defined in the common sharing project and are imported through dependencies while corresponding XML files are in the respective system projects of the microservices. Seckillcloud, for instance, experienced the problem. Because our program analyzes each microservice individually, the aforementioned project structure will result in analysis errors. The data extracted for the above-mentioned issues was rectified during the manual inspection phase.

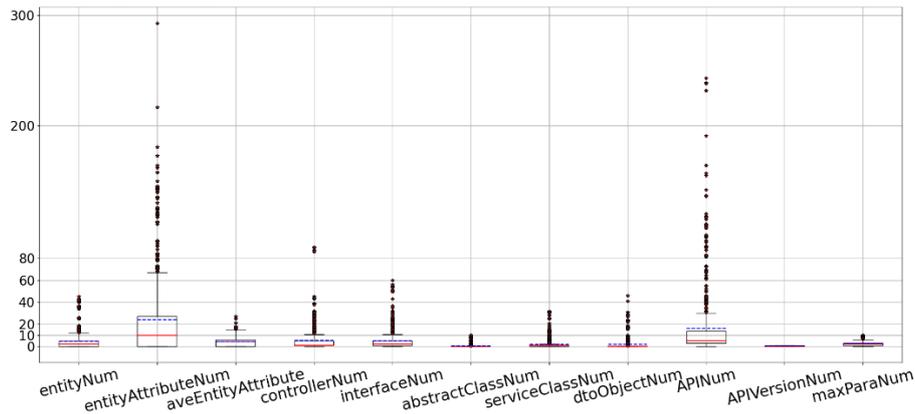

**Fig. 4.** Distribution of data points in each metric.

Lastly, we can give a simple example of using this dataset. Nano Service mentioned by Tighilt et.al. [14], which means overly fine-grained service granularity, can be evaluated comprehensively by considering metrics such as the number of entity classes, the number of controller classes, and the number of APIs exposed in the microservices. We can label the dataset with the bad smell in design and apply machine learning, heuristic algorithms, or other means to validate the effectiveness of the dataset being labeled and compare the efficiency of these different detection means and get the better one.

## 6 Threats to validity

The data in this paper is derived from open-source microservice systems. The limitation of our search criteria may have led to the exclusion of exceptional open-source microservice systems. Moreover, during the collection process, we observed that some open-source microservice systems are primarily developed for demonstration purposes. It means that they cannot represent the whole open-source ecosystem and the scale, and the scale and maturity of these collected microservice systems differ significantly compared to the large-scale micro-service systems used by Internet companies in the real world. This implies that the extracted data would require validation for its effectiveness if used for machine learning purposes.

Furthermore, the challenges encountered in Section IV also highlight the variations in technology and development standards among microservice systems developed by different teams. Accomplishing 100% accurate automatic extraction poses a difficult challenge, given the diverse external technologies and development standards employed in microservice systems. The effectiveness of the extraction program should be validated for various external technologies and development standards, along with its external validity in the Spring Cloud style microservice systems applied in the real world. Meanwhile, the detection of entity classes for NoSQL was not considered, indicating a limitation in the research.

Nevertheless, this paper believes that the collected microservice systems can, to some extent, embody the structure and development standards of Spring Boot style microservice. The extraction program has also shown good effectiveness in application, and this work holds practical significance.

## 7        Conclusions and future work

In this paper, we have collected 55 Spring Cloud style microservice systems based on Maven from GitHub, forming a catalog. To intuitively evaluate the granularity of microservices, their design, and the interactions between different microservices, we devised extraction metrics based on microservices with the Spring Boot three-tier architecture and developed an extraction program. Based on the collected microservice systems, the established metrics were extracted, and manual verification was performed, resulting in a feature dataset derived from open-source microservice systems. This dataset can contribute to the application of machine learning and heuristic algorithms in microservice bad smell research, as well as provide a foundational sample dataset for comparing the efficiency of different microservice bad smell detection methods. We believe that this dataset will facilitate further research in the domain of microservice bad smells.

Our future work will focus on expanding the catalog of microservice systems, continuously modifying, and improving the feature extraction program, and expanding the dataset. Additionally, attempts will be made to label the dataset for specific sets of microservice bad smells and apply machine learning algorithms to research in the domain of microservice bad smells. The curation of a catalog of microservice bad smells that can be detected using the feature dataset presented in this paper is underway.

**Table 2.** Introduction of metrics.

| Name | Introduction | Classification |
|---|---|---|
| codeSize | The effective lines of source code in all .java files of the microservice | Code lines |
| entityNum | The number of object classes used for persistent storage in the microservice. | |
| entityAttributeNum | The number of attributes contained in the microservice entity classes. | |
| controllerNum | The number of controllers in the microservice. | |
| interfaceNum | The number of interfaces in the microservice. | |
| abstractClassNum | The number of abstract classes in the microservice. | |
| serviceClassNum | The number of service implementation classes in the microservice. | Classes |
| dtoClassNum | The number of Data Transfer Object classes in the microservice. | |
| APINum | The number of APIs exposed by the microservice. | |
| maxParaNum | The maximum value of the parameter list size of all APIs exposed. | |
| APIVersionSet | The collection of versions for APIs of the microservice. It's a set of Strings representing versions. | APIs |
| serviceImplCall | The data structure is a map, where the keys correspond to the methods of service implementation classes invoked in the controller classes within the microservice. The values in the collection represent the count of internal invocations for each method. | |
| serviceCall | The data structure is $Map <serviceA, <ServiceB,time>>$, where ServiceA acts as the primary key, ServiceB serves as the nested key, and the associated value, times, denotes the frequency of ServiceA invoking ServiceB. | Interaction |
| serviceCalled | The data structure is $Map <serviceA, <ServiceB,time>>$, where ServiceA acts as the primary key, ServiceB serves as the nested key, and the associated value, times, denotes the frequency of ServiceA being invoked by ServiceB. | |
| aveEntityAttribute | The ratio obtained by dividing the total number of attributes in entity classes by the total number of entity classes. | Classes |
| APIVersionNum | The size of APIVersionSet. | APIs |
| serviceImplCallNum | The total number of times the controller layer invokes methods within the service implementation layer. | |
| maxServiceCall | The maximum number of times the current microservice invokes other microservices. | |
| serviceCallGate | The number of distinct microservices invoked by the current microservice. | |
| serviceCallPer | The percentage of distinct microservices invoked by the current microservice out of the total number of microservices in the system. | Interaction |
| maxServiceCalled | The maximum number of times the current microservice is called by other microservices. | |
| serviceCalledGate | The number of distinct microservices that invoke the current microservice. | |
| serviceCalledPer | The percentage of distinct microservices that invoke the current microservice out of the total number of microservices in the system. | |

**Table 3.** Display of System Data Compared with Manual Verification.

| System | Data size | EN | EAN | CN | IN | AN | SN | DTON | APIN | MPN | SIC | SC | SCD |
|---|---|---|---|---|---|---|---|---|---|---|---|---|---|
| gpmall | 10 | 30% | 30% | 100% | 100% | 100% | 100% | 100% | 100% | 100% | 100% | 100% | 100% |
| mogu_blog_v2 | 42 | 100% | 100% | 100% | 100% | 100% | 100% | 100% | 100% | 100% | 100% | 100% | 100% |
| mall4cloud | 55 | 100% | 100% | 100% | 100% | 100% | 100% | 100% | 100% | 100% | 100% | 90.90% | 90.90% |
| Scblogs | 18 | 100% | 83.33% | 100% | 100% | 100% | 100% | 100% | 100% | 100% | 100% | 100% | 100% |
| Seckillcloud | 20 | 66.67% | 66.67% | 100% | 100% | 100% | 100% | 100% | 100% | 100% | 100% | 85.71% | 71.43% |
| spring-pet-clinic-micro-services | 66 | 100% | 100% | 100% | 100% | 100% | 100% | 100% | 100% | 100% | 100% | 100% | 100% |
| train-tickets | 338 | 100% | 100% | 100% | 100% | 100% | 100% | 100% | 100% | 100% | 100% | 100% | 100% |

EN: entityNum. EAN: entityAttributeNum. CN: controllerNum. IN: interfaceNum. AN: abstractClassNum. SN: serviceClassNum. DTON: dtoClassNum. APIN: APINum. MPN: maxParamNum. SIC: serviceImplCall. SCS: serviceCall. SCD: serviceCalled.
Data Size: the number of data points extracted from the microservice system.